\newcommand{\CLASSINPUTtoptextmargin}{0.85in}
\newcommand{\CLASSINPUTbottomtextmargin}{0.97in}
\documentclass[conference]{IEEEtran}
\IEEEoverridecommandlockouts
\usepackage{standalone}%

\usepackage{cite}
\usepackage{amsmath,amssymb,amsfonts}
\usepackage{mathtools}
\usepackage{bm}
\usepackage{algorithmic}
\usepackage{graphicx}
\usepackage{textcomp}
\usepackage{subcaption}
\usepackage{xcolor}
\def\BibTeX{{\rm B\kern-.05em{\sc i\kern-.025em b}\kern-.08em T\kern-.1667em\lower.7ex\hbox{E}\kern-.125emX}}

\usepackage{trfsigns}
\usepackage{placeins}
\graphicspath{{inkscape/}} %
\usepackage{tabularx}         
\usepackage{multirow}  
\usepackage{hhline}                    
\usepackage[shortcuts,acronym]{glossaries}
\usepackage{blkarray, bigstrut}

\usepackage{pgfplots}
\pgfplotsset{compat=newest} %
\usepackage{tikz}
\usetikzlibrary{calc}
\pgfkeys{
    /includestandaloneresized/.is family, /includestandaloneresized,
    default/.style = {
        width = \textwidth,
        ratio = 0.61803398875},
    width/.estore in = \itgWidth,
    ratio/.estore in = \itgRatio
}
\newlength\figurewidth
\newlength\figureheight

\def\unit{0.78cm}  %
\usetikzlibrary{shapes.geometric, positioning, calc, shapes.arrows, backgrounds}
\usepackage{amsmath,amssymb,amsfonts}
\usetikzlibrary{shapes, shapes.geometric, positioning, calc, shapes.arrows, backgrounds, shapes.gates.logic.US, shapes.gates.logic.IEC}

\makeatletter
\def\ps@IEEEtitlepagestyle{%
    \def\@oddfoot{\mycopyrightnotice}%
    \def\@evenfoot{}%
}
\def\mycopyrightnotice{%
    {\footnotesize
        This work has been submitted to the IEEE for possible publication. Copyright may be transferred without notice, after which this version may no longer be accessible.\hfill}%
    \gdef\mycopyrightnotice{}%
}

\IEEEoverridecommandlockouts\IEEEpubid{\makebox[\columnwidth]{\hfill} \hspace{\columnsep}\makebox[\columnwidth]{ }}
\begin{document}
\setlength\abovecaptionskip{0.45\baselineskip}
\setlength{\textfloatsep}{0.35\baselineskip}
\def\@IEEEfigurecaptionsepspace{\vskip\abovecaptionskip\relax}%
\def\@IEEEtablecaptionsepspace{\vskip\abovecaptionskip\relax}%
\title{Uniform vs. Non-Uniform Coarse Quantization in\\Mutual Information Maximizing LDPC Decoding}
\IEEEaftertitletext{\vspace{-0.5\baselineskip}} %
\author{
\IEEEauthorblockN{Philipp Mohr, Gerhard Bauch}
\IEEEauthorblockA{\textit{Hamburg University of Technology} \\
\textit{Institute of Communications}\\
21073 Hamburg, Germany \\
Email: \{philipp.mohr, bauch\}@tuhh.de}
}
\maketitle

\newacronym{app}{APP}{a-posteriori probability}
\newacronym{bp}{BP}{belief propagation}
\newacronym{llr}{LLR}{log-likelihood ratio}
\newacronym{lut}{LUT}{lookup table}
\newacronym{luts}{LUTs}{lookup tables}
\newacronym{ib}{IB}{information bottleneck}
\newacronym{ldpc}{LDPC}{low-density parity-check}
\newacronym{qc}{QC}{quasi-cyclic}
\newacronym{omsq}{OMSQ}{quantized offset-min-sum}
\newacronym[longplural={variable nodes}]{vn}{VN}{variable node}
\newacronym[longplural={check nodes}]{cn}{CN}{check node}

\begin{abstract}
%auto-ignore
Recently, low-resolution LDPC decoders have been introduced that perform mutual information maximizing signal processing. 
However, the optimal quantization in variable and check nodes requires expensive non-uniform operations. 
Instead, we propose to use uniform quantization with a simple hardware structure, which reduces the complexity of individual node operations approximately by half and shortens the decoding delay significantly. 
Our analysis shows that the loss of preserved mutual information resulting from restriction to uniform quantization is very small. 
Furthermore, the error rate simulations with regular LDPC codes confirm that the uniform quantization causes only minor performance degradation within 0.01 dB compared to the non-uniform alternative.
Due to the complexity reduction, especially the proposed 3-bit decoder is a promising candidate to replace 4-bit conventional decoders.
\end{abstract}

%auto-ignore
\newcommand\myvec{\boldsymbol}%
\newcommand\mymat[1]{\boldsymbol{\mathbf{#1}}}
\newcommand\myup{\mathrm}
\newcommand\mysamplespace{\mathcal}
\newcommand\mycard[1]{\lvert #1\rvert}
\newcommand\myset[1]{\left\{#1\right\}}
\newcommand\mynotset[1]{\sim\{#1\}}
\newcommand\mylabel{\mathrm}
\newcommand\myvar[1]{\mathsf{#1}}
\FloatBarrier
\section{Introduction}
Low-density parity-check codes (LDPC) are established in many modern communication applications, like fiber-optic, Ethernet, wireless or NAND flash systems. 
In particular, their capacity-approaching error correction capabilities and the existence of iterative message-passing decoding algorithms with high parallelism make LDPC codes an excellent choice. But the constantly increasing demand for high data rates and energy-efficient systems requires continuous improvements.

To alleviate the complexity caused by the message transfers between variable and check nodes, finite alphabet decoders that maximize the preserved mutual information have been proposed. Excellent performance is achieved while using only a few bits for the exchanged messages\cite{kurkoski_noise_2008, lewandowsky_information-optimum_2018, meidlinger_quantized_2015, he_mutual_2019,stark_machine_2021, mohr_coarsely_2021,wang_rec_2022}.
Conventional algorithms, like the offset or normalized min-sum decoders\cite{jinghu_chen_reduced-complexity_2005}, work with higher resolutions to achieve similar performance.

However, most of the mutual information maximizing decoders use non-uniform operations in the node updates that can cause higher hardware complexity than in the conventional algorithms.
One option to perform mutual information maximizing node updates is the two-input lookup table decomposition technique\cite{lewandowsky_information-optimum_2018,kurkoski_noise_2008}. Each lookup table encodes a non-uniform compression mapping for any realization of the two input messages, which maximizes the preserved mutual information between the output message and a relevant variable. 
Although this avoids exponentially increasing table sizes,  the multiple compression steps in each node update introduce performance degradation\cite{he_mutual_2019, mohr_coarsely_2021}.
To reduce this loss, we decided to focus on the so-called computational domain approach\cite{he_mutual_2019,lee_memory-efficient_2005}. The technique is characterized by three steps: First, small single-input lookup tables translate the low-resolution messages to higher-resolution representation values. Secondly, arithmetic operations combine those values into one high-resolution extrinsic message. 
Finally, the extrinsic message is compressed in a single step with non-uniform threshold quantization which enables optimal preservation of relevant information in the compressed output message\cite{kurkoski_quantization_2014}. 
As we will show, the non-uniform threshold quantization consumes large amounts of the hardware resources dedicated to check and variable nodes. 
In particular, those resources include multiple high-resolution comparison operations and memory for the threshold values.

In this paper, we propose to reduce the complexity by restricting to symmetric uniform quantization with equally spaced thresholds. 
With that restriction, the outputs of the translation tables can be scaled such that the uniform quantization consists only of a simple bit-shift operation. 
In this way, no high-resolution comparators or memory for the thresholds are required.
Furthermore, the optimization complexity is potentially reduced since the design involves fewer degrees of freedom than the non-uniform quantization. We observe only a small loss in the preserved mutual information and minor performance degradation in terms of error rates. The main contributions are summarized as follows.
\begin{itemize}
    \item The non-uniform quantization in the computational domain approach is replaced with uniform quantization for check and variable nodes. The proposed design eliminates the complexity caused by non-uniform quantization.
    \item A variable node structure is developed leading to symmetric probability distributions of the exchanged messages. Hence, memory requirements are reduced by half and translation tables avoided when using the minimum approximation in the check node.
    \item The non-uniform and uniform solutions are compared in detail with respect to the threshold levels, effect on probability distributions and preserved mutual information. The discrete density evolution method confirms that uniform quantization introduces only a small loss in terms of mutual information.
    \item Error rates simulations reveal that uniform quantization achieves performance within 0.01 dB compared to optimal non-uniform quantization. 
\end{itemize}

\section{Mutual Information Maximizing Decoders}
We assume a binary LDPC code with parity check matrix $\mymat{H}\in \{0,1\}^{N_c\times N}$ which can be represented by a Tanner graph with $N$ \glspl{vn} and $N_c$ \glspl{cn}. The encoder maps the information bits $\myvec{u}{\in}\{0,1\}^{K\times 1}$ to code bits $\myvec{b}{\in}\{0,1\}^{N\times 1}$ satisfying $\mymat{H}\myvec{b}{=}\myvec{0}$. 
For the transmission, we consider binary phase-shift keying (BPSK) symbols which are disturbed by additive white Gaussian noise (AWGN) at the receiver.
A mutual information maximizing symmetric channel quantizer maps the received symbols to messages $\myvec{t}^{ch}{\in}\mysamplespace{T}^{N}$ with finite alphabet $\mysamplespace{T}$ of bit width $w$\cite{lewandowsky_information-optimum_2018}. 
In the decoder, $w$-bit messages are exchanged over multiple decoding iterations between variable and check nodes. We apply a flooding schedule, where one decoding iteration consists of updating all check nodes and, subsequently, all variable nodes. 
However, the proposed techniques can be applied also to other schedules\cite{mohr_coarsely_2021}. In the first iteration the channel messages are directly forwarded to the corresponding check nodes.

\subsection{Computational Domain Node Updates}\label{sec:comp_domain_updates}
\begin{figure}[t]
    \centering
        \begin{tikzpicture} [scale=1.0, x=\unit,y=\unit, node distance=1.0*\unit and 1.0*\unit]   
    \node(sgnsum)[rectangle, draw=black, fill=white, minimum width = 4*\unit]{$\oplus$};
    
    \node(m1)[above=of sgnsum.170, yshift=1*\unit]{$t^v_1$};
    \node(m2)[above=of sgnsum.150, yshift=1*\unit]{$t^v_2$};
    \node(mdc)[above=of sgnsum.10, yshift=1*\unit]{$t^v_{d_c}$};
    \node (mdots) at ($(m2)!0.5!(mdc)$) {$\dots$};
    
    \draw[-latex](m1.south)--(sgnsum.170);
    \draw[-latex](m2.south)--(sgnsum.150);
    \draw[-latex](mdc.south)--(sgnsum.10);
    
    \node(sgnminus1)[rectangle, draw=black, below=of sgnsum.170, yshift=0.0*\unit, inner sep = 2pt]{$\oplus$};
    \node(sgnminus2)[rectangle, draw=black, below=of sgnsum.150, yshift=0.0*\unit, inner sep = 2pt]{$\oplus$};
    \node(sgnminusdc)[rectangle, draw=black, below=of sgnsum.10, yshift=0.0*\unit, inner sep = 2pt]{$\oplus$};
    
    \begin{scope}[on background layer]
        \draw[-latex](m1.south)|-+(0.7, -1.7)|-(sgnminus1.east);
        \draw[-latex](m2.south)|-+(0.7, -1.7)|-(sgnminus2.east);
        \draw[-latex](mdc.south)|-+(0.7, -1.7)|-(sgnminusdc.east);
    \end{scope}
    
    \draw[-latex](sgnsum.south)--+(0,-0.15)-|(sgnminus1);
    \draw[-latex](sgnsum.south)--+(0,-0.15)-|(sgnminus2);
    \draw[-latex](sgnsum.south)--+(0,-0.15)-|(sgnminusdc);
    
    \node(mc1)[below=of sgnminus1, yshift=-1.1*\unit]{$t^c_1$};
    \node(mc2)[below=of sgnminus2, yshift=-1.1*\unit]{$t^c_2$};
    \node(mcdc)[below=of sgnminusdc, yshift=-1.1*\unit]{$t^c_{d_c}$};
    \node (mcdots) at ($(mc2)!0.5!(mcdc)$) {$\dots$};
    
    \draw[-latex](sgnminus1)--(mc1.north);
    \draw[-latex](sgnminus2)--(mc2.north);
    \draw[-latex](sgnminusdc)--(mcdc.north);
    
    \node(abssum)[rectangle, draw=black, fill=white, right=of sgnsum, xshift=-0.25*\unit, minimum width = 4*\unit]{$+$};

    \node(absphim1)[rectangle, draw=black, above=of abssum.170, yshift=-0.5*\unit]{$|\phi_\Delta^v|$};
    \node(absphim2)[rectangle, draw=black, above=of abssum.150, yshift=-0.5*\unit]{$|\phi_\Delta^v|$};
    \node(absphimdc)[rectangle, draw=black, above=of abssum.10, yshift=-0.5*\unit]{$|\phi_{\Delta}^v|$};
   
    \draw[-latex](m1.south)--+(0,-0.4)-|(absphim1);
    \draw[-latex](m2.south)--+(0,-0.3)-|(absphim2);
    \draw[-latex](mdc.south)--+(0,-0.2)-|(absphimdc);
    
    \draw[-latex](absphim1)--(absphim1|-abssum.north);
    \draw[-latex](absphim2)--(absphim2|-abssum.north);
    \draw[-latex](absphimdc)--(absphimdc|-abssum.north);
    
    \node(absminus1)[rectangle, draw=black, below=of abssum.170, yshift=0.0*\unit, inner sep = 2pt]{$+$};
    \node(absminus2)[rectangle, draw=black, below=of abssum.150, yshift=0.0*\unit, inner sep = 2pt]{$+$};
    \node(absminusdc)[rectangle, draw=black, below=of abssum.10, yshift=0.0*\unit, inner sep = 2pt]{$+$};
    
    \begin{scope}[on background layer]
        \draw[-latex](absphim1.south)|-+(0.7, -0.15)|-(absminus1.east)node[xshift=0.15*\unit, yshift=0.15*\unit] {-};
        \draw[-latex](absphim2.south)|-+(0.7, -0.15)|-(absminus2.east)node[xshift=0.15*\unit, yshift=0.15*\unit] {-};
        \draw[-latex](absphimdc.south)|-+(0.7, -0.15)|-(absminusdc.east)node[xshift=0.15*\unit, yshift=0.15*\unit] {-};
    \end{scope}
    
    \draw[-latex](abssum.south)--+(0,-0.15)-|(absminus1);
    \draw[-latex](abssum.south)--+(0,-0.15)-|(absminus2);
    \draw[-latex](abssum.south)--+(0,-0.15)-|(absminusdc);
    
    \node(quant1)[rectangle, draw=black, below=of absminus1, yshift=0.5*\unit]{$Q$};
    \node(quant2)[rectangle, draw=black, below=of absminus2, yshift=0.5*\unit]{$Q$};
    \node(quantdc)[rectangle, draw=black, below=of absminusdc, yshift=0.5*\unit]{$Q$};
   
    \draw[-latex](absminus1)--(quant1);
    \draw[-latex](absminus2)--(quant2);
    \draw[-latex](absminusdc)--(quantdc);
    
    \draw[-latex](quant1.south)--+(0, -0.4)-|(mc1.north);
    \draw[-latex](quant2.south)--+(0, -0.5)-|(mc2.north);
    \draw[-latex](quantdc.south)--+(0, -0.6)-|(mcdc.north);
   
   \node (labelquant1input) at (quant1)[xshift=0.3*\unit, yshift=0.6*\unit, red]{$|y|$};
   \node (labelquant1output) at (quant1)[xshift=0.3*\unit, yshift=-0.6*\unit, red]{$|t|$};
  \node (labely1sgn) at (sgnminus1)[xshift=0.55*\unit, yshift=-0.6*\unit]{\footnotesize$\operatorname{sgn}(y)$};
   \node (labelt1sgn) at (sgnminus1|-quant1)[xshift=0.55*\unit, yshift=-0.6*\unit]{\footnotesize$\operatorname{sgn}(t)$};
    \end{tikzpicture}
    \vspace{-0.2cm}
    \caption{A full check node update.}
    \label{fig:full_cnu_hardware}
\end{figure}
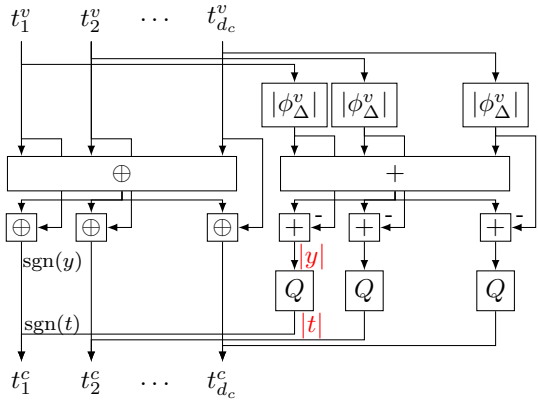
As explained in the introduction, in this work we focus on reducing the complexity of the computational domain approaches for check and variable nodes\cite{lee_memory-efficient_2005,he_mutual_2019}. In the following, we extend this framework by replacing the non-uniform with uniform quantization. 
\subsubsection{Check Node}
A check node of degree $d_c$ receives the variable node messages $t^v_i\in \mysamplespace{T}$ with $i{\in }\mathcal{N}_c{=}\{1,\ldots,d_c\}$. Each message $t_i^v$ provides information about a code bit $x_i^c$ that participates in the parity check equation $x^c_{1}{\oplus}x^c_{2} {\oplus} \ldots {\oplus} x^c_{d_c}{=}0$. %
The mutual information maximizing node update (derivation in appendix \ref{app:derivation_cn_operation}) yields\cite{he_mutual_2019}
\begin{align}
\begin{split}
t^c_j=Q^c\left(\prod_{i\in \mathcal{N}_c\setminus \{j\}}\operatorname{sgn}(\phi^v_i(t^v_i))\sum_{i\in \mathcal{N}_c\setminus \{j\}} |\phi^v_i(t^v_i)|\right).
\label{equ:cn_update}
\end{split}
\end{align}
In (\ref{equ:cn_update}), $\phi^{v}_i$ are small translation tables computed according to
 $\phi^{v}_i(t^v_i) = \operatorname{sgn}(L(x^c_i|t^v_i))(-\log|\tanh L(x^c_i|t^v_i)/2|)$ with the log-likelihood ratio (LLR) being $L(x|t)=p(x{=}0|t)/p(x{=}1|t)$.
The quantizer $Q^c$ is defined by a threshold set %
and $\max_{Q^c} I(\myvar{X}^c_j;\myvar{T}_j^c)$ can be performed, e.g., using the sequential information bottleneck algorithm\cite{lewandowsky_information-optimum_2018}. %
\subsubsection{Variable Node}
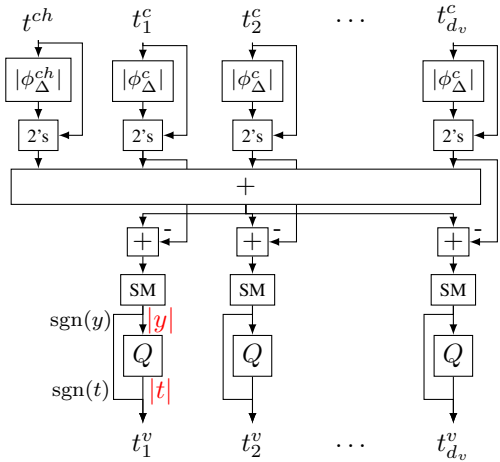
\begin{figure}[t]
    \centering
        \begin{tikzpicture} [scale=1.0, x=\unit,y=1.0*\unit, node distance=0.3*\unit and 1.0*\unit]   
    \node(sum)[rectangle, draw=black, fill=white, minimum width = 8*\unit, minimum height = 0.1*\unit]{$+$};
    
    \node(twoch)[rectangle, draw=black, above=of sum.175, yshift=0*\unit]{\scriptsize 2's};
    \node(two1)[rectangle, draw=black, above=of sum.170, yshift=0*\unit]{\scriptsize 2's};
    \node(two2)[rectangle, draw=black, above=of sum.70, yshift=0*\unit]{\scriptsize 2's};
    \node(twodv)[rectangle, draw=black, above=of sum.5, yshift=0*\unit]{\scriptsize 2's};
    
    \node(absphimch)[rectangle, draw=black, above=of twoch, yshift=0.0*\unit]{\footnotesize$|\phi_\Delta^{ch}|$};
    \node(absphim1)[rectangle, draw=black, above=of two1, yshift=0.0*\unit]{\footnotesize$|\phi_\Delta^c|$};
    \node(absphim2)[rectangle, draw=black, above=of two2, yshift=0.0*\unit]{\footnotesize$|\phi_\Delta^c|$};
    \node(absphimdv)[rectangle, draw=black, above=of twodv, yshift=0.0*\unit]{\footnotesize$|\phi_\Delta^c|$};
    
    \node(mch)[above=of absphimch, yshift=0*\unit]{$t^{ch}$};
    \node(m1)[above=of absphim1, yshift=0*\unit]{$t^c_1$};
    \node(m2)[above=of absphim2, yshift=0*\unit]{$t^c_{2}$};
    \node(mdv)[above=of absphimdv, yshift=0*\unit]{$t^c_{d_v}$};
    \node (mdots) at ($(m2)!0.5!(mdv)$) {$\dots$};
    
    \node(minus1)[rectangle, draw=black, below=of two1, yshift=-1*\unit, inner sep = 2pt]{$+$};
    \node(minus2)[rectangle, draw=black, below=of two2, yshift=-1*\unit,inner sep = 2pt]{$+$};
    \node(minusdv)[rectangle, draw=black, below=of twodv, yshift=-1*\unit, inner sep = 2pt]{$+$};
    
    \node(sm1)[rectangle, draw=black, below=of minus1, yshift=0*\unit]{\scriptsize SM};
    \node(sm2)[rectangle, draw=black, below=of minus2, yshift=0*\unit]{\scriptsize SM};
    \node(smdv)[rectangle, draw=black, below=of minusdv, yshift=0*\unit]{\scriptsize SM};
    
    \node(quant1)[rectangle, draw=black, below=of sm1, yshift=-0.2*\unit]{$Q$};
    \node(quant2)[rectangle, draw=black, below=of sm2, yshift=-0.2*\unit]{$Q$};
    \node(quantdv)[rectangle, draw=black, below=of smdv, yshift=-0.2*\unit]{$Q$};
    
    \node(mv1)[below=of quant1, yshift=-0.5*\unit]{$t^v_1$};
    \node(mv2)[below=of quant2, yshift=-0.5*\unit]{$t^v_2$};
    \node(mvdv)[below=of quantdv, yshift=-0.5*\unit]{$t^v_{d_v}$};
    \node (mvdots) at ($(mv2)!0.5!(mvdv)$) {$\dots$};
    
    \draw[-latex](mch.south)--(absphimch);
    \draw[-latex](absphimch.south)--(absphimch|-twoch.north);
    \draw[-latex](mch.south)|-+(0.75,-0.05)|-(twoch.east);
    \draw[-latex](twoch.south)--(twoch|-sum.north);
    
    \draw[-latex](m1.south)--(absphim1);
    \draw[-latex](absphim1.south)--(absphim1|-two1.north);
    \draw[-latex](m1.south)|-+(0.75,-0.05)|-(two1.east);
    \draw[-latex](two1.south)--(two1|-sum.north);
    \draw[-latex](sum.south)--+(0,-0.15)-|(minus1.north);
    \begin{scope}[on background layer]
    \draw[-latex](two1.south)|-+(0.75, -0.15)|-(minus1.east)node[xshift=0.15*\unit, yshift=0.15*\unit] {-};
    \end{scope}
    \draw[-latex](minus1.south)--(sm1.north);
    \draw[-latex](sm1.south)--(quant1.north);
    \draw[-latex](quant1.south)--(mv1.north);
    \draw[-latex](sm1.south)|-+(-0.5,-0.15)|-+(0,-1.6)-|(mv1.north);
    
    \draw[-latex](m2.south)--(absphim2);
    \draw[-latex](absphim2.south)--(absphim2|-two2.north);
    \draw[-latex](m2.south)|-+(0.75,-0.05)|-(two2.east);
    \draw[-latex](two2.south)--(two2|-sum.north);
    \draw[-latex](sum.south)--+(0,-0.15)-|(minus2.north);
    \begin{scope}[on background layer]
    \draw[-latex](two2.south)|-+(0.75, -0.15)|-(minus2.east)node[xshift=0.15*\unit, yshift=0.15*\unit] {-};
    \end{scope}
    \draw[-latex](minus2.south)--(sm2.north);
    \draw[-latex](sm2.south)--(quant2.north);
    \draw[-latex](quant2.south)--(mv2.north);
    \draw[-latex](sm2.south)|-+(-0.5,-0.15)|-+(0,-1.6)-|(mv2.north);
    
    \draw[-latex](mdv.south)--(absphimdv);
    \draw[-latex](absphimdv.south)--(absphimdv|-twodv.north);
    \draw[-latex](mdv.south)|-+(0.75,-0.05)|-(twodv.east);
    \draw[-latex](twodv.south)--(twodv|-sum.north);
    \draw[-latex](sum.south)--+(0,-0.15)-|(minusdv.north);
    \begin{scope}[on background layer]
    \draw[-latex](twodv.south)|-+(0.75, -0.15)|-(minusdv.east)node[xshift=0.15*\unit, yshift=0.15*\unit] {-};
    \end{scope}
    \draw[-latex](minusdv.south)--(smdv.north);
    \draw[-latex](smdv.south)--(quantdv.north);
    \draw[-latex](quantdv.south)--(mvdv.north);
    \draw[-latex](smdv.south)|-+(-0.5,-0.15)|-+(0,-1.6)-|(mvdv.north);

    \node (labelquant1input) at (quant1)[xshift=0.3*\unit, yshift=0.6*\unit, red]{$|y|$};
    \node (labelquant1input) at (quant1)[xshift=-1.05*\unit, yshift=0.6*\unit]{\footnotesize$\operatorname{sgn}(y)$};
    \node (labelquant1output) at (quant1)[xshift=0.3*\unit, yshift=-0.6*\unit, red]{$|t|$};
    \node (labelquant1output) at (quant1)[xshift=-1.05*\unit, yshift=-0.6*\unit]{\footnotesize$\operatorname{sgn}(t)$};
    
    \end{tikzpicture}
    \vspace{-0.2cm}
    \caption{A full variable node update. %
    }
    \label{fig:full_vnu_hardware}
\end{figure}
\begin{figure}
     \vspace{-0.3cm}
    \begin{subfigure}[b]{0.47\linewidth}
            \begin{tikzpicture} [scale=1.0, x=\unit,y=\unit, node distance=0.3*\unit and 0.5*\unit]     
    \node(b3)[rectangle, draw=black]{$\tau_3$};
    \node(comp1)[rectangle, draw=black, below=of b3]{$\le$};
    
    \node(b1)[rectangle, draw=black, below=of comp1, xshift=-0.35*\unit]{$\tau_1$};
    \node(b5)[rectangle, draw=black, below=of comp1, xshift=0.35*\unit]{$\tau_5$};
    \node(comp2)[rectangle, draw=black, below=of comp1, yshift=-1*\unit]{$\le$};
    
    \node(b0)[rectangle, draw=black, below=of comp2, xshift=-1.05*\unit]{$\tau_0$};
    \node(b2)[rectangle, draw=black, below=of comp2, xshift=-0.35*\unit]{$\tau_2$};
    \node(b4)[rectangle, draw=black, below=of comp2, xshift=0.35*\unit]{$\tau_4$};
    \node(b6)[rectangle, draw=black, below=of comp2, xshift=1.05*\unit]{$\tau_6$};
    
    \node(comp3)[rectangle, draw=black, below=of comp2, yshift=-1*\unit]{$\le$};
    
    \node(y)[left=of comp1, xshift=-1.2*\unit, label={[xshift=0.3*\unit, yshift=-0.15*\unit, red]{$|y|$}}]{};
    \node(t)[right=of comp3, xshift=1.5*\unit, label={[xshift=-0.3*\unit, yshift=-0.15*\unit,red]{$|t|$}}]{};
    
    \draw[-latex](b3.south)--(comp1.north);
    \draw[-latex](b1.south-|comp2.north)--(comp2.north);
    \draw[-latex](b0.south-|comp3.north)--(comp3.north);
    
    \draw[-latex](comp1.south)--(comp1.south|-b1.north);
    \draw[-latex](comp2.south)--(comp2.south|-b0.north);
    
    \draw[-latex](y.east)--(comp1.west);
    \draw[-latex](y.east)--+(0.5,0)|-(comp2.west);
    \draw[-latex](y.east)--+(0.5,0)|-(comp3.west);
    
    \draw[-latex](comp1.east)--+(1.3,0)|-(t.west);
    \draw[-latex](comp2.east)--+(1.3,0)|-(t.west);
    \draw[-latex](comp3.east)--(t.west);

    \coordinate (tbitwidth) at ($(t.west)+(-0.4*\unit, 0)$);
    \draw ($(tbitwidth)+(0, 0.1*\unit)$)--($(tbitwidth)-(0, 0.1*\unit)$)node[yshift=-0.15*\unit]{\scriptsize$3$};
    
    \coordinate (ybitwidth0) at ($(y.east)+(0.2*\unit, 0)$);
    \draw ($(ybitwidth0)+(0, 0.1*\unit)$)--($(ybitwidth0)-(0, 0.1*\unit)$)node[yshift=-0.15*\unit]{\scriptsize$8$};
    
    \coordinate (ybitwidth1) at ($(comp1.west)+(-0.8*\unit, 0)$);
    \draw ($(ybitwidth1)+(0, 0.1*\unit)$)--($(ybitwidth1)-(0, 0.1*\unit)$)node[yshift=-0.15*\unit]{\scriptsize$8$};
    
    \coordinate (ybitwidth2) at ($(comp2.west)+(-0.8*\unit, 0)$);
    \draw ($(ybitwidth2)+(0, 0.1*\unit)$)--($(ybitwidth2)-(0, 0.1*\unit)$)node[yshift=-0.15*\unit]{\scriptsize$8$};
    
    \coordinate (ybitwidth3) at ($(comp3.west)+(-0.8*\unit, 0)$);
    \draw ($(ybitwidth3)+(0, 0.1*\unit)$)--($(ybitwidth3)-(0, 0.1*\unit)$)node[yshift=-0.15*\unit]{\scriptsize$8$};
    
    \coordinate (comp1outbitwidth) at ($(comp1.east)+(1*\unit, 0)$);
    \draw ($(comp1outbitwidth)+(0, 0.1*\unit)$)--($(comp1outbitwidth)-(0, 0.1*\unit)$)node[yshift=-0.15*\unit]{\scriptsize$1$};
    
    \coordinate (comp2outbitwidth) at ($(comp2.east)+(1*\unit, 0)$);
    \draw ($(comp2outbitwidth)+(0, 0.1*\unit)$)--($(comp2outbitwidth)-(0, 0.1*\unit)$)node[yshift=-0.15*\unit]{\scriptsize$1$};
    
    \coordinate (comp3outbitwidth) at ($(comp3.east)+(1*\unit, 0)$);
    \draw ($(comp3outbitwidth)+(0, 0.1*\unit)$)--($(comp3outbitwidth)-(0, 0.1*\unit)$)node[yshift=-0.15*\unit]{\scriptsize$1$};
     
    \end{tikzpicture}
        \caption{Non-uniform}
        \label{fig:hardware_quant_nonuniform}
    \end{subfigure}
    \hfill
    \begin{subfigure}[b]{0.5\linewidth}
            \begin{tikzpicture} [scale=1.0, x=\unit,y=\unit, node distance=0.5*\unit and 0.5*\unit]   
     \node(y)[yshift=1*\unit, label={[xshift=0.3*\unit, yshift=-0.15*\unit, red]{$|y|$}}]{};
     \node(orclip)[or gate US, draw, logic gate inputs=nnn, right=of y, xshift=0.9*\unit, label={[xshift=-0.0*\unit, yshift=-1.4*\unit]{OR}}]{};
     
     \node(orclip1)[or gate US, draw, logic gate inputs=nn, right=of y, xshift=2.7*\unit, yshift=-1*\unit]{};
     \node(orclip2)[or gate US, draw, logic gate inputs=nn, below=of orclip1]{};
     \node(orclip3)[or gate US, draw, logic gate inputs=nn, below=of orclip2]{};
     
     \node(deadend1)[circle,fill=black,inner sep=0pt,minimum size=4pt, below=of orclip, yshift=-3*\unit]{};
     \node(deadend2)[circle,fill=black,inner sep=0pt,minimum size=4pt, below=of deadend1]{};
     
     \node(t)[right=of orclip3, xshift=0.2*\unit, label={[xshift=-0.3*\unit, yshift=-0.15*\unit, red]{$|t|$}}]{};
     
     \draw[-latex](y.east)-|+(0.5,0)|-(orclip.input 1);
     \draw[-latex](y.east)-|+(0.5,0)|-(orclip.input 2);
     \draw[-latex](y.east)-|+(0.5,0)|-(orclip.input 3);
     
     \draw[-latex](orclip.east)-|+(0.3,0)|-(orclip1.input 1);
     \draw[-latex](y.east)-|+(0.5,0)|-(orclip1.input 2);
     
     \draw[-latex](orclip.east)-|+(0.3,0)|-(orclip2.input 1);
     \draw[-latex](y.east)-|+(0.5,0)|-(orclip2.input 2);
     
     \draw[-latex](orclip.east)-|+(0.3,0)|-(orclip3.input 1);
     \draw[-latex](y.east)-|+(0.5,0)|-(orclip3.input 2);
     
     \draw[-latex](y.east)-|+(0.5,0)|-(deadend1);
     \draw[-latex](y.east)-|+(0.5,0)|-(deadend2);
     
     \draw[-latex](orclip1.east)-|+(0.15,0)|-(t.west);
     \draw[-latex](orclip2.east)-|+(0.15,0)|-(t.west);
     \draw[-latex](orclip3.east)-|+(0.15,0)|-(t.west);
     
     \coordinate (ybitwidth0) at ($(y.east)+(0.2*\unit, 0)$);
     \draw ($(ybitwidth0)+(0, 0.1*\unit)$)--($(ybitwidth0)-(0, 0.1*\unit)$)node[yshift=-0.15*\unit]{\scriptsize$8$};
     
     \coordinate (tbitwidth) at ($(t.west)-(0.3*\unit, 0)$);
     \draw ($(tbitwidth)+(0, 0.1*\unit)$)--($(tbitwidth)-(0, 0.1*\unit)$)node[yshift=-0.15*\unit]{\scriptsize$3$};
     
     \node (MSB) at ($(orclip.input 1)+(-0.5*\unit, 0.3*\unit)$){\footnotesize MSB};
     \node (LSB) at ($(deadend2.west-|MSB)+(0, 0.3*\unit)$){\footnotesize LSB};
     
     \node (rshift) at ($(deadend1)+(1*\unit, -0.3*\unit)$){$r=2$};

    \end{tikzpicture}
        \caption{Uniform}
        \label{fig:hardware_quant_uniform}
    \end{subfigure}
    \caption{Hardware schematics for performing quantization with $w_y=9$ bit as the bit width of $y$ and $w=4$ bit for $t$.}
    \label{fig:hardware_quant}
\end{figure}
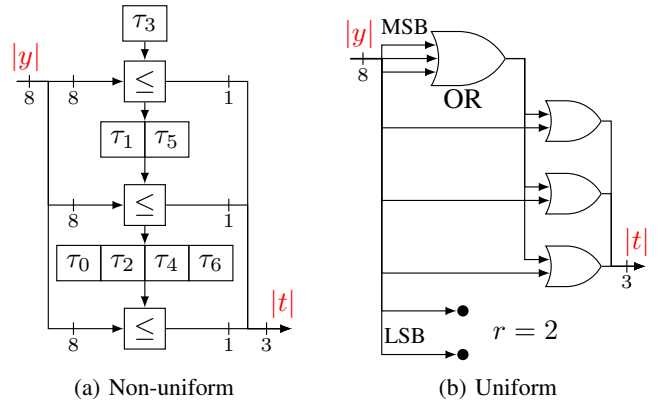
A variable node of degree $d_v$ receives the channel and check node messages, $t^{ch}\in \mysamplespace{T}$ and $t_i^c\in \mysamplespace{T}$ with $i{\in}\mathcal{N}^v{=}\{1,\ldots, d_v\}$. All messages provide information about the underlying code bit $x^v$. The mutual information maximizing variable node update yields\cite{he_mutual_2019, stark_machine_2021, mohr_coarsely_2021}:
\begin{align}
    t^v_j = Q^v\left(\phi^{ch}(t^{ch}) + \sum_{i\in \mathcal{N}_v\setminus \{j\}} \phi^c_i(t^c_i)\right),
    \label{equ:vn_update}
\end{align}
where $\phi^{ch}(t^{ch}){=}L(t^{ch}|x^v)$ and $\phi^c_i(t^c_i){=}L(t^c_i|x^v)$ represent small translation tables. 
Similar to the check node, the quantizer $Q^v$ is optimized according to $\max_{Q^v} I(\myvar{X}^v;\myvar{T}^v_j)$.

\subsection{An Efficient Hardware Design with Uniform Quantization}\label{sec:hardware}
In practice, the internal resolution after translation is limited to $w_{\phi}$ bits. Computations in the check node update (\ref{equ:cn_update}) and variable node update (\ref{equ:vn_update}) are performed with integer instead of real valued numbers. Therefore, the translation lookup table $\phi{=}\phi^v, \phi^{ch}\text{ or }\phi^c$ is scaled to an integer range with
\begin{align}
    \phi_{\Delta}(t){=}\operatorname{sgn}(\phi(t)) \operatorname{min}\left(\left\lfloor\frac{1}{\Delta}|\phi(t)|{+}\frac{1}{2}\right\rfloor{,} 2^{w_{\phi}{-}1}{-}1\right)
\end{align}
where $\phi_{\Delta}(t)\approx \frac{1}{\Delta}\phi(t)$ and $\Delta\in \mathbb{R}^+$.

Fig.~\ref{fig:full_cnu_hardware} and Fig.~\ref{fig:full_vnu_hardware} depict possible hardware structures to perform the update for all outputs of the check node or variable node, respectively. For an efficient implementation we assume symmetric distributions of the input messages with $p(\myvar{X}{=}0, \myvar{T}{=}t)=p(\myvar{X}{=}1,\myvar{T}{=}{-}t)$ when using a sign-magnitude (SM) format with $t\in\mysamplespace{T}{=}\{{-}2^{w-1},\ldots,{-}1,{+}1,\ldots,{+}2^{w-1}\}$. 
In this way, translation tables only require the message's magnitude bits to perform the mapping into the computational domain, which cuts the memory demand by half. 
\subsubsection{Symmetry Preserving Variable Node Update}
In contrast to the check node, the variable node performs signed addition using the 2's complement format, such that the same hardware component can deal with addition and subtraction. 
Therefore, after translation, one conversion step from SM to the 2's complement format is required, which can be implemented with a few logic gates. After summation, another conversion back to the SM format is done. Hence, only the magnitude part must be quantized. Note, that this can cause an asymmetric distribution $p(x^v,y)$: If the summation yields $y_{\text{2's}}{=}0$, conversion to the SM format leads to $y{=}{+}1$ and, therefore, $p(\myvar{Y}{=}{+}1)>p(\myvar{Y}{=}{-}1)$. 
For canceling out asymmetric distributions in the design phase, we use two types of variable nodes: One adds up negative LLRs and inverts the sign at the output; the other requires no modifications. In this way, we enforce a symmetric distribution 
\begin{align}
p(x^v,y){=}p(x^v,\myvar{Y}_\text{2's}{=}y){+}\begin{dcases}p(x^v,\myvar{Y}_\text{2's}{=}0)/2 & y{=}\pm 1\\
0 & y\neq \pm 1
\end{dcases}.
\end{align}
\subsubsection{Non-Uniform Quantization}
We perform the symmetric non-uniform quantization with $Q$ as $Q^v$ or $Q^c$ according to
\begin{align}
Q(y)=\operatorname{sgn}(y)\begin{dcases}
1 & |y|{<}\tau_{0}\\
i & \tau_{i}{<}|y|{<}\tau_{i+1}, 0{<}i{<}2^{w-1}{-}2\\
2^{w-1} & |y|{<}\tau_{2^{w-1}-2} 
\end{dcases}.
\end{align}
The corresponding hardware implementation is depicted in Fig.~\ref{fig:hardware_quant_nonuniform}. The threshold quantization with a binary search technique involves $w{-}1$ high resolution comparisons. For the full update in total $d_v(w{-}1)$ comparisons are required, which can cause even more complexity than the addition operations. Furthermore, memory and logic are necessary to hold and select the appropriate thresholds $\{\tau_0,\ldots, \tau_{2^{w-1}-2}\}$.

\subsubsection{Uniform Quantization}\label{sec:uniform_quantization}
We perform symmetric uniform quantization with $Q$ as $Q^v$ or $Q^c$ according to %
\begin{align}
Q(y)=\operatorname{sgn}(y)\min\left(\lfloor |y|/2^r\rfloor+1,2^{w-1}\right).
\end{align}
An efficient hardware implementation is proposed in Fig.~\ref{fig:hardware_quant_uniform} with a right-shift operation by $r\in \mathbb{N}_0$ positions and a subsequent clipping such that the desired resolution $w$ is obtained. By modifying $r$ and the scaling factor $\Delta$ for the translation tables, any uniform boundary spacing, $\Delta(\tau_{i+1}{-}\tau_i){=}\Delta 2^{r}$ w.r.t. the reconstructed LLRs $\Delta{\cdot} y\approx L(x|y)$, can be achieved.
The optimal uniform quantization is obtained with a grid based search aiming for $\max_{\Delta, r} I(\myvar{X};\myvar{T})$ where $\myvar{T}{\coloneqq} \myvar{T}^v_j$ ($\myvar{T}^c_j$) and $\myvar{X}{\coloneqq} \myvar{X}^v$ ($\myvar{X}^c_j$) for the variable (check) node. Only $(w_y{-}w{-}r){+}(w{-}1)$ logic OR gates are required, which is negligible complexity compared to the hardware used for the non-uniform threshold quantization in Fig.~\ref{fig:hardware_quant_nonuniform}. 
For the check node, an offset $\kappa$ in $Q^c(y')$ with $y'{=}\operatorname{sgn}(y)(|y|+\kappa)$ (moves first boundary $\tau_0$ closer to the decision threshold), can improve the performance especially for higher iterations.

\subsection{Check Node with Minimum Approximation}
Instead of (\ref{equ:cn_update}), another option is the check node update with minimum approximation from \cite{meidlinger_quantized_2015},
\begin{align}
\begin{split}
t^c_j=\prod_{i\in \mathcal{N}_c\setminus \{j\}}\operatorname{sgn}(t^v_i)\min_{i\in \mathcal{N}_c\setminus \{j\}} |t^v_i|,
\label{equ:cn_update_min}
\end{split}
\end{align}
where the sign magnitude format and symmetric design of Section~\ref{sec:hardware} is assumed. The number of comparisons required in hardware can be reduced to $d_c{+}\lceil \log_2 d_c \rceil-2$\cite{lee_low-complexity_2015}.

\subsection{Complexity Evaluation}
\begin{table}
    \centering
    \caption{Complexity of variable and check nodes.}%
    \label{table:complexity_vnu}
    \setlength{\tabcolsep}{3pt}
    \begin{tabular}{ccccc}
        &\multirow{2}{*}{\shortstack[c]{node\\variant}}& \multirow{2}{*}{\shortstack[c]{additions/\\comparisons}}  & \multirow{2}{*}{\shortstack[c]{translation\\count $\phi$}} & \multirow{2}{*}{\shortstack[c]{memory usage\\ { in bit}}} \\
        &&&&\\
        \hline
        \multirow{4}{*}{CN}
        &non-uniform &$(w{+}1)d_c{-}2$                   & $d_c$                 &$(w_\phi{+}w_s{-}1)2^{w-1}$\\
        &uniform &$2d_c{-}2$                         & $d_c$                 &$w_\phi2^{w-1}$          \\
        &min. approx.&$d_c{+}\lceil \log_2 d_c \rceil-2$                           &-                      &-               \\
        &OMSQ    &$d_c{+}\lceil \log_2 d_c \rceil$   &-                      &-               \\
        \hline
        \multirow{3}{*}{VN}     
        &non-uniform &$(w{+}1)d_v{-}1$                   & $d_v{+}1$             &$(2w_\phi{+}w_s{-}3)2^{w-1}$\\
        &uniform &$2d_v{-}1$                         & $d_v{+}1$             &$(2w_\phi{-}2)2^{w{-}1}$\\
        &OMSQ    &$2d_v{-}1$                         &-                      &-\\
    \end{tabular}
    \vspace{-0.0cm}
\end{table}
In Table \ref{table:complexity_vnu}, the check and variable node complexity of the proposed decoders is compared to the offset min-sum (OMSQ) algorithm\cite{jinghu_chen_reduced-complexity_2005}. The non-uniform decoder involves the highest number of operations for check and variable nodes, due to the more expensive quantization. Also, the memory requirements are highest to represent the translation tables and boundaries. In contrast, the uniform decoder can avoid comparisons and memory for boundaries. Further complexity reduction is achieved by using the minimum approximation in the check node, where only a first and second minimum search must be performed and no translation tables are required\cite{jinghu_chen_reduced-complexity_2005, meidlinger_quantized_2015}. Note, that the two-minima search involves additional multiplexers in hardware, which are not considered in Table~\ref{table:complexity_vnu}. The OMSQ check node has slightly higher complexity due to the offset operation.

\section{Performance Evaluation}
\subsection{Density Evolution Analysis of Different Quantizers}\label{sec:dde_analysis_different_quantizers}
\begin{figure}[t]
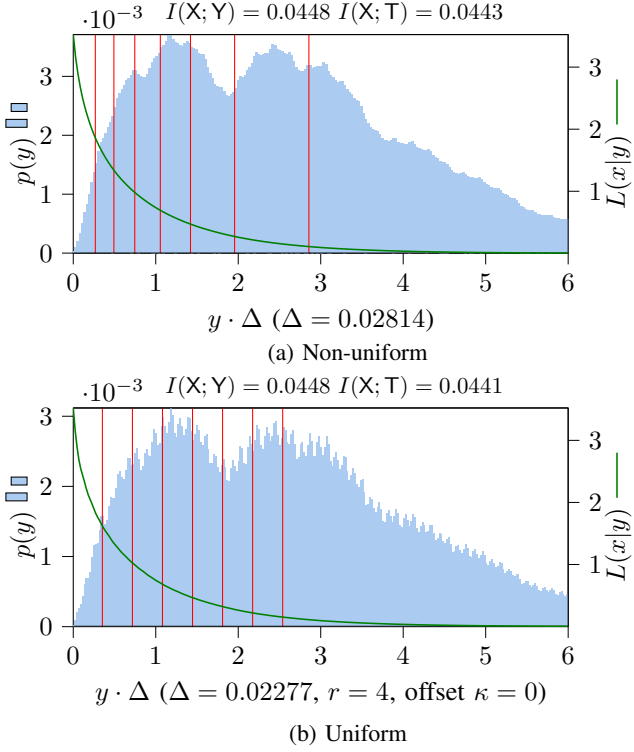

    \begin{subfigure}[b]{0.5\textwidth}
        \hspace{-1.0cm}
        
    \pgfkeys{/includestandaloneresized, default, width=0.9\linewidth, ratio=0.55}
    \setlength\figurewidth{\itgWidth}
    \setlength\figureheight{\itgRatio\figurewidth}
    \includestandalone{standalones/cn_nonuniform/source}

        \vspace{-0.2cm}
        \caption{Non-uniform}
        \label{fig:cn_nonuniform}
    \end{subfigure}
    \hfill
    \begin{subfigure}[b]{0.5\textwidth}
        \vspace{-0.0cm}
        \hspace{-1.0cm}
        
    \pgfkeys{/includestandaloneresized, default, width=0.9\linewidth, ratio=0.55}
    \setlength\figurewidth{\itgWidth}
    \setlength\figureheight{\itgRatio\figurewidth}
    \includestandalone{standalones/cn_uniform/source}

        \vspace{-0.1cm}
        \caption{Uniform}
        \label{fig:cn_uniform}
    \end{subfigure}
    \caption{Check node distributions and boundary placements.}
    \label{fig:cn_comparisons}
    \vspace{-0.0cm}
\end{figure}
\begin{figure}[t]
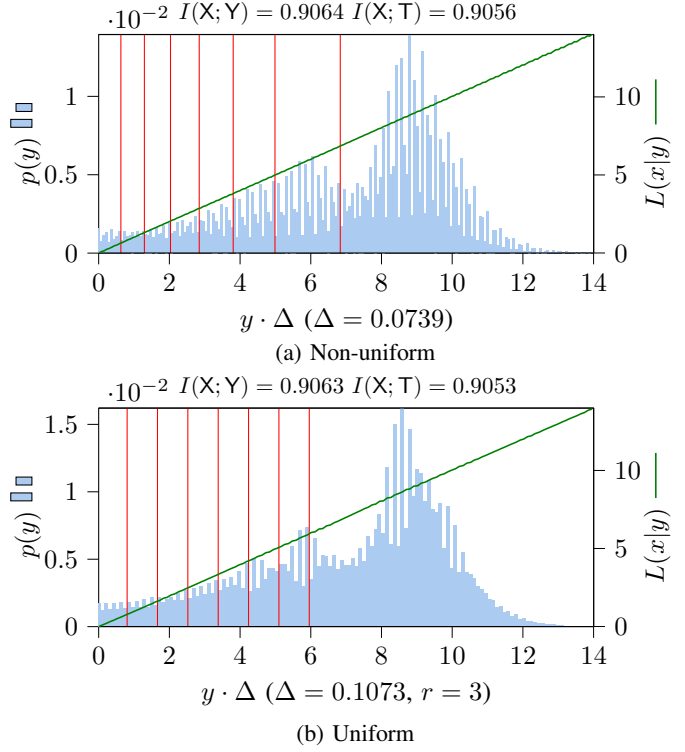

    \begin{subfigure}[b]{0.5\textwidth}
        \hspace{-1.0cm}
        
    \pgfkeys{/includestandaloneresized, default, width=0.9\linewidth, ratio=0.55}
    \setlength\figurewidth{\itgWidth}
    \setlength\figureheight{\itgRatio\figurewidth}
    \includestandalone{standalones/vn_nonuniform/source}

        \vspace{-0.2cm}
        \caption{Non-uniform}
        \label{fig:vn_nonuniform}
    \end{subfigure}
    \begin{subfigure}[b]{0.5\textwidth}
        \vspace{-0.0cm}
        \hspace{-1.0cm}
        
    \pgfkeys{/includestandaloneresized, default, width=0.9\linewidth, ratio=0.55}
    \setlength\figurewidth{\itgWidth}
    \setlength\figureheight{\itgRatio\figurewidth}
    \includestandalone{standalones/vn_uniform/source}

        \vspace{-0.1cm}
        \caption{Uniform}
        \label{fig:vn_uniform}
    \end{subfigure}
    \caption{Variable node distributions and boundary placements.}
    \vspace{-0.0cm}
    \label{fig:vn_comparisons}
\end{figure}
For the mutual information maximizing decoder design, discrete density evolution\cite{kurkoski_noise_2008} is used to track the distributions of messages with respect to relevant variables. %

In Fig.~\ref{fig:cn_comparisons} and Fig.~\ref{fig:vn_comparisons}, the optimization results and corresponding distributions are shown for check and variable node designs in the first iteration. The degree distribution corresponds to the code of Section~\ref{sec:scenario1} with $d_c{=}32$ and $d_v{=}6$. The design SNR is set to $E_b/N_0{=}3.3$ dB. The internal resolution is $w_\phi=8$\,bit.

In the following analysis we consider the quantizer operation of (\ref{equ:cn_update}) and (\ref{equ:vn_update}) as information bottleneck setups, where $\myvar{X}$, $\myvar{Y}$ and $\myvar{T}$ are defined as the relevant, observed and compressed variable, respectively\cite{lewandowsky_information-optimum_2018}.

For the check node $\myvar{X}{=}\myvar{X}^{c}_j$ and $\myvar{T}{=}\myvar{T}^c_j$. For the variable node $\myvar{X}{=}\myvar{X}^{v}$ and $\myvar{T}{=}\myvar{T}^v_j$. In both cases, the observed variable $\myvar{Y}$ is the quantizer input.

Please note, that the horizontal axis depicts only half of the sample space $\mathcal{Y}$ w.r.t. positive LLRs, since the other half is symmetric. Furthermore, the integers $y$ are scaled with $\Delta$ to facilitate comparisons of the boundary placements in Fig.~\ref{fig:cn_comparisons} and \ref{fig:vn_comparisons}. %

\subsubsection{Check Node}
Looking at Fig.~\ref{fig:cn_nonuniform}, it can be seen that the boundaries are more dense in regions where $L(x|y)$ has steeper slope. This is reasonable, since the optimization procedure aims at minimizing the loss of relevant information
\begin{align}
\begin{split}
     &\min_{Q}\small I(\myvar{X};\myvar{Y}){-}I(\myvar{X};\myvar{T})\\
    =&\min_{Q} \sum_y p(y)D_{\text{KL}}(p(x|y)||p(x|t{=}Q(y))),
    \label{equ:mi_diff}
\end{split}
\end{align}
where the Kullback Leibler divergence is given by $D_{\text{KL}}(p(x|y)||p(x|t))=\sum_x p(x|y)\log \frac{p(x|y)}{p(x|t)}$. From (\ref{equ:mi_diff}) we can conclude that the loss is increased more severely, if $D_{\text{KL}}$ (always non-negative) is large and if the event $y$ has high probability of occurrence $p(y)$. Consequently, the boundaries in Fig.~\ref{fig:cn_nonuniform} are more dense in regions where the meaning $L(x|y)$ changes rapidly, to combat large divergences. %

In Fig.~\ref{fig:cn_uniform}, we observe that the uniform quantization mimics the behavior of the non-uniform solution. The mutual information difference between non-uniform and uniform check node ($0.0443{-}0.0441{=}0.0003$) is negligible compared to the difference between non-uniform and minimum approximation check node ($0.0443{-}0.0407{=}0.0036$).  Also note, that $p(y)$ depends on the spacing $\Delta{=}0.02277$ (uniform) which is different from $\Delta{=}0.02814$ (non-uniform).
\subsubsection{Variable Node}
For evaluation in Fig.~\ref{fig:vn_comparisons} we use the same check node implementation (minimum approximation) to ensure equal input distributions.
For the non-uniform quantizer at the variable node in Fig.~\ref{fig:vn_nonuniform}, the rate of change in $L(x|y)$ is constant w.r.t. to $y$. Therefore, the boundary placement is determined primarily by the fact that $D_\text{KL}$ has less contribution for large $L(x|y)$, and, secondly, by $p(y)$. Especially the first effect causes the optimization to place the boundaries more densely close to $y=0$.

The uniform quantization at the variable node in Fig.~\ref{fig:vn_uniform} turns out to be very effective, too. Compared to Fig.~\ref{fig:vn_nonuniform} the difference in mutual information $0.9056-0.9053=0.0003$ is similar to the loss observed for the uniform check node.
\subsubsection{Summary}
Overall, we can conclude that the boundary placement depends mainly on three effects:
\begin{enumerate}
    \item Clustering of events that are close in meaning, e.g. merging the events $y_1$ and $y_2$ where $L(x|y_1)\approx L(x|y_2)$, causes only minor relevant information loss in (\ref{equ:mi_diff}).
    \item Clustering in regions with low reliability, i.e., where $|L(x|y)|$ is small, leads to more information loss than clustering in regions with high reliability. This effect is less dominant than the first one, which becomes apparent by carefully comparing the boundary placement w.r.t. $L(x|y)$ of the check node and the variable node results.
    \item Clustering in regions where $p(y)$ is small, has minor influence on the relevant information loss in (\ref{equ:mi_diff}) and, therefore, the boundaries are less dense in those regions.
\end{enumerate}

\subsection{Decoder Design and Error Rate Performance}
In this section we compare the performance in terms of mutual information and error rates. All scenarios use either 3 or 4 bits for the channel quantizer, variable node and check node messages. The internal resolution is $w_{\phi}=8$\,bit. Smaller resolutions like $w_{\phi}{=}6$\,bit introduced degradation of about 0.01\,dB. Table \ref{table:setups_vns} summarizes the node configurations used in this work. The first part of each decoder label specifies the check node and the second part the variable node. In the error rate simulations, the design $E_b/N_0$ is optimized for best frame or bit error rate performance.

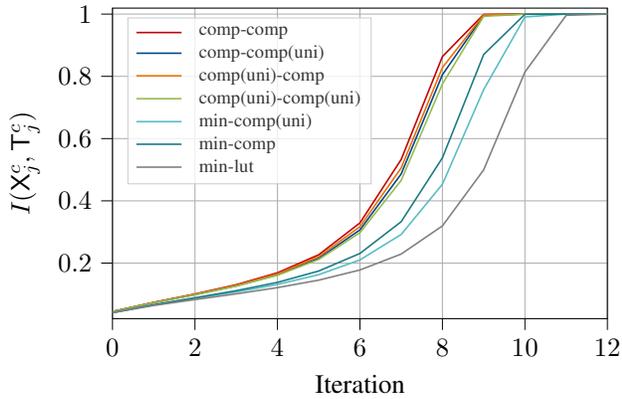
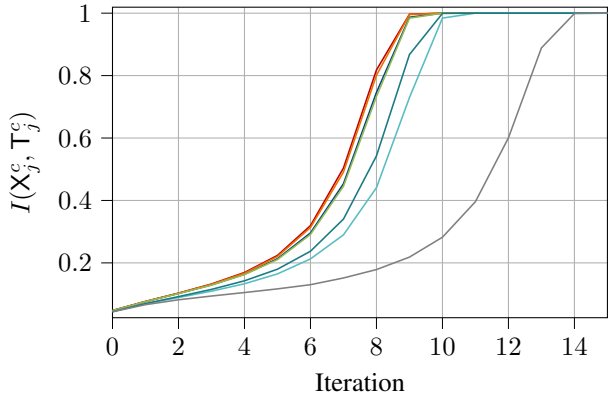
\begin{figure}[t]
    \begin{subfigure}[b]{0.5\textwidth}
        
    \pgfkeys{/includestandaloneresized, default, width=0.9\textwidth, ratio=0.7}
    \setlength\figurewidth{\itgWidth}
    \setlength\figureheight{\itgRatio\figurewidth}
    \pgfplotsset{every axis/.append style={
        scaled y ticks = false, 
        scaled x ticks = false, 
        y tick label style={/pgf/number format/fixed,
                            /pgf/number format/precision=5},
        x tick label style={/pgf/number format/fixed,
                            /pgf/number format/precision=5}
    }
}
        
\pgfplotsset{legend style={nodes={scale=0.7, transform shape}}}
\pgfplotsset{ every non boxed x axis/.append style={x axis line style=-},
               every non boxed y axis/.append style={y axis line style=-}}
\begin{tikzpicture}

\definecolor{color0}{rgb}{0,0.270588235294118,0.541176470588235}
\definecolor{color1}{rgb}{0.925490196078431,0.431372549019608,0}
\definecolor{color2}{rgb}{0.584313725490196,0.729411764705882,0.309803921568627}
\definecolor{color3}{rgb}{0.341176470588235,0.741176470588235,0.764705882352941}
\definecolor{color4}{rgb}{0.0980392156862745,0.47843137254902,0.517647058823529}

\begin{axis}[
height=\figureheight,
legend cell align={left},
legend style={
  fill opacity=0.8,
  draw opacity=1,
  text opacity=1,
  at={(0.03,0.97)},
  anchor=north west,
  draw=white!80!black
},
tick align=outside,
tick pos=left,
width=\figurewidth,
x grid style={white!69.0196078431373!black},
xlabel={Iteration},
xmajorgrids,
xmin=0, xmax=12,
xminorgrids,
xtick style={color=black},
y grid style={white!69.0196078431373!black},
ylabel={$I(\mathsf{X}^c_j, \mathsf{T}^c_j)$},
ymajorgrids,
ymin=0.0214831479956826, ymax=1.01918660494126,
yminorgrids,
ytick style={color=black}
]
\addplot [semithick, red!75.2941176470588!black]
table {%
0 0.0442942044508879
1 0.0737700899604763
2 0.100700804299723
3 0.130434701700222
4 0.168818460479292
5 0.226749175138493
6 0.329080214913793
7 0.532876541099152
8 0.863596001489565
9 0.999039580805577
10 1
11 1
12 1
};
\addlegendentry{comp-comp}
\addplot [semithick, color0]
table {%
0 0.0442942044508879
1 0.0729275256152623
2 0.0987364884593316
3 0.126834962728468
4 0.162647802268192
5 0.215265256167736
6 0.306638079687023
7 0.486378326201748
8 0.805211029850743
9 0.995825455399553
10 0.999999998011311
11 1
12 1
};
\addlegendentry{comp-comp(uni)}
\addplot [semithick, color1]
table {%
0 0.0440816856983793
1 0.0731890235027036
2 0.0995962994596951
3 0.12842283392223
4 0.165536304763746
5 0.220782298130532
6 0.317009107839652
7 0.505769308050952
8 0.828248409328293
9 0.997475797277366
10 1
11 1
12 1
};
\addlegendentry{comp(uni)-comp}
\addplot [semithick, color2]
table {%
0 0.0440816856983793
1 0.0724605196783029
2 0.0981815929457399
3 0.125854931957491
4 0.160857464983386
5 0.211734457125231
6 0.297758128241468
7 0.465657334403977
8 0.775555258961015
9 0.99274784829517
10 0.999999993895446
11 1
12 1
};
\addlegendentry{comp(uni)-comp(uni)}
\addplot [semithick, color3]
table {%
0 0.0406697529369437
1 0.0651622995312846
2 0.0861221432594722
3 0.107024559273964
4 0.13140566826163
5 0.162809766302465
6 0.210163381101374
7 0.291974164771986
8 0.453271899097742
9 0.757377386375687
10 0.99064343820741
11 0.999999993558915
12 1
};
\addlegendentry{min-comp(uni)}
\addplot [semithick, color4]
table {%
0 0.0406697529369437
1 0.0663704265288773
2 0.0886275758263633
3 0.111322634343939
4 0.137983119899819
5 0.174474228567368
6 0.230998530587419
7 0.332937942781553
8 0.538092979061738
9 0.869961122190981
10 0.999339062225712
11 1
12 1
};
\addlegendentry{min-comp}
\addplot [semithick, gray]
table {%
0 0.0406697529369437
1 0.0638113787305839
2 0.0827098354087673
3 0.101190346700143
4 0.121168010875033
5 0.145173755630167
6 0.177865882759737
7 0.229015804728532
8 0.319475739728497
9 0.499553422611626
10 0.813694260805563
11 0.996048593259138
12 0.999999995324545
};
\addlegendentry{min-lut}
\end{axis}

\end{tikzpicture}

        \caption{4 bits, $E_b/N_0=3.3$ dB}
        \label{fig:mi_ctv_code2_4bit}
    \end{subfigure}
    \hfill
    \begin{subfigure}[b]{0.5\textwidth}
        
    \pgfkeys{/includestandaloneresized, default, width=0.9\textwidth, ratio=0.7}
    \setlength\figurewidth{\itgWidth}
    \setlength\figureheight{\itgRatio\figurewidth}
    \pgfplotsset{every axis/.append style={
        scaled y ticks = false, 
        scaled x ticks = false, 
        y tick label style={/pgf/number format/fixed,
                            /pgf/number format/precision=5},
        x tick label style={/pgf/number format/fixed,
                            /pgf/number format/precision=5}
    }
}
        
\pgfplotsset{legend style={nodes={scale=0.7, transform shape}}}
\pgfplotsset{ every non boxed x axis/.append style={x axis line style=-},
               every non boxed y axis/.append style={y axis line style=-}}
\begin{tikzpicture}

\definecolor{color0}{rgb}{0,0.270588235294118,0.541176470588235}
\definecolor{color1}{rgb}{0.925490196078431,0.431372549019608,0}
\definecolor{color2}{rgb}{0.584313725490196,0.729411764705882,0.309803921568627}
\definecolor{color3}{rgb}{0.341176470588235,0.741176470588235,0.764705882352941}
\definecolor{color4}{rgb}{0.0980392156862745,0.47843137254902,0.517647058823529}

\begin{axis}[
height=\figureheight,
tick align=outside,
tick pos=left,
width=\figurewidth,
x grid style={white!69.0196078431373!black},
xlabel={Iteration},
xmajorgrids,
xmin=0, xmax=15,
xminorgrids,
xtick style={color=black},
y grid style={white!69.0196078431373!black},
ylabel={$I(\mathsf{X}^c_j, \mathsf{T}^c_j)$},
ymajorgrids,
ymin=0.024361918995922, ymax=1.01913015845106,
yminorgrids,
ytick style={color=black}
]
\addplot [semithick, red!75.2941176470588!black]
table {%
0 0.0473168640698253
1 0.0765151156751376
2 0.103145106561671
3 0.132335955463817
4 0.169235561773952
5 0.223895695511204
6 0.319360758614524
7 0.503890663442613
8 0.817903071375144
9 0.996489671823962
10 0.999999998793211
11 1
12 1
13 1
14 1
15 1
};
\addplot [semithick, color0]
table {%
0 0.0473168640698253
1 0.0759725603450427
2 0.102116783547036
3 0.129442122468295
4 0.16348601220963
5 0.213204101411737
6 0.295903442505106
7 0.453477753172387
8 0.745858375412024
9 0.986883506546064
10 0.999999941856412
11 1
12 1
13 1
14 1
15 1
};
\addplot [semithick, color1]
table {%
0 0.047185820188258
1 0.0762672297766165
2 0.102608787178061
3 0.131149046252612
4 0.166905610597421
5 0.220378239412556
6 0.313275795964001
7 0.490970607695791
8 0.800911600912733
9 0.995434635048143
10 1
11 1
12 1
13 1
14 1
15 1
};
\addplot [semithick, color2]
table {%
0 0.047185820188258
1 0.075680303609068
2 0.101629843784054
3 0.128789940575192
4 0.162534394761337
5 0.210583717653452
6 0.291110390021382
7 0.445438735416294
8 0.732909441666843
9 0.983940719045674
10 0.99999988438904
11 1
12 1
13 1
14 1
15 1
};
\addplot [semithick, color3]
table {%
0 0.0434920774469823
1 0.0689685958103154
2 0.0890718120418756
3 0.109344814062907
4 0.133104418244185
5 0.164690708729834
6 0.212703906335817
7 0.289752420592288
8 0.440822824668907
9 0.729527831558214
10 0.984268208546588
11 0.999999891108763
12 1
13 1
14 1
15 1
};
\addplot [semithick, color4]
table {%
0 0.0434920774469823
1 0.0697340426021478
2 0.0919309156093535
3 0.115070798198801
4 0.142540200382454
5 0.179758707056519
6 0.23728105575349
7 0.340267849625744
8 0.542625156257428
9 0.867364723407795
10 0.999194156581313
11 1
12 1
13 1
14 1
15 1
};
\addplot [semithick, gray]
table {%
0 0.0434920774469823
1 0.066108138715066
2 0.0817054027091591
3 0.0938236455688472
4 0.104977437824931
5 0.116716560831844
6 0.130131957817499
7 0.151643131965954
8 0.178607459093336
9 0.218478974307613
10 0.282147058188806
11 0.396102064537117
12 0.600626422566832
13 0.888702965973236
14 0.998738949184756
15 0.999999996830039
};
\end{axis}

\end{tikzpicture}

        \caption{3 bits, $E_b/N_0=3.4$ dB}
        \label{fig:mi_ctv_code2_3bit}
    \end{subfigure}
    \caption{Evolution of mutual information for scenario 1.}
    \label{fig:evolution_of_mi}
\end{figure}

\begin{table}[t]
    \centering
    \caption{Check and variable node variants.}%
    \label{table:setups_vns}
    \setlength{\tabcolsep}{3pt}
    \begin{tabular}{ccccc}
        node&label & description & equation & reference\\ 
        \hline
        \multirow{3}{*}{CN}&comp      &  non-uniform comp. domain & (\ref{equ:cn_update}) & \cite{he_mutual_2019}\\ 
        &comp(uni)         &  uniform comp. domain &(\ref{equ:cn_update}), \ref{sec:uniform_quantization} & proposed\\
        &min               &  minimum approximation & (\ref{equ:cn_update_min}) & \cite{meidlinger_quantized_2015}\\
        \hline
        \multirow{3}{*}{VN}&comp      &  non-uniform comp. domain&(\ref{equ:vn_update}) & \cite{he_mutual_2019}\\    
        &comp(uni)         &  uniform comp. domain & (\ref{equ:vn_update}), \ref{sec:uniform_quantization} & proposed\\
        &lut              &  two-input lookup tables & - &\cite{lewandowsky_information-optimum_2018}
    \end{tabular}
    \vspace{-0.0cm}
\end{table}

\begin{figure}[t]
    \vspace{-0.3cm}
    \hspace*{-0.7cm}   %
    
    \pgfkeys{/includestandaloneresized, default, width=0.95\linewidth, ratio=1.0}
    \setlength\figurewidth{\itgWidth}
    \setlength\figureheight{\itgRatio\figurewidth}
    \pgfplotsset{legend style={nodes={scale=0.7, transform shape}}}
\pgfplotsset{ every non boxed x axis/.append style={x axis line style=-},
               every non boxed y axis/.append style={y axis line style=-}}
\begin{tikzpicture}

\definecolor{color0}{rgb}{0,0.270588235294118,0.541176470588235}
\definecolor{color1}{rgb}{0.584313725490196,0.729411764705882,0.309803921568627}
\definecolor{color2}{rgb}{0.341176470588235,0.741176470588235,0.764705882352941}

\begin{axis}[
height=\figureheight,
legend cell align={left},
legend style={
  fill opacity=0.8,
  draw opacity=1,
  text opacity=1,
  at={(0.03,0.03)},
  anchor=south west,
  draw=white!80!black,
  at={(0.01,0.01)}
},
log basis y={10},
tick align=outside,
tick pos=left,
width=\figurewidth,
x grid style={white!69.0196078431373!black},
xlabel={\(\displaystyle E_b/N_0\) in dB},
xmajorgrids,
xmin=3.85, xmax=4.5,
xminorgrids,
xtick style={color=black},
y grid style={white!69.0196078431373!black},
ylabel={Frame Error Rate},
ymajorgrids,
ymin=1e-06, ymax=0.01,
yminorgrids,
ymode=log,
ytick style={color=black}
]
\addplot [semithick, black,  densely dashed, mark=*, mark size=1.5, mark options={solid}]
table {%
2 1
2.1 1
2.2 1
2.3 1
2.4 1
2.5 0.9995
2.6 0.9995
2.7 0.999
2.8 0.9905
2.9 0.976
3 0.9265
3.1 0.8725
3.2 0.739
3.3 0.6225
3.4 0.414
3.5 0.26125
3.6 0.150571428571429
3.7 0.075
3.8 0.0281111111111111
3.9 0.010234693877551
4 0.00296165191740413
4.1 0.000763941940412529
4.2 0.000175377060680463
4.3 3.39453593286643e-05
4.4 5.1150895140665e-06
4.5 8.20660960337456e-07
};
\addlegendentry{OMSQ 4-bit}
\addplot [semithick, red!75.2941176470588!black, mark=square*, mark size=1.225, mark options={solid}]
table {%
2 1
2.1 1
2.2 1
2.3 1
2.4 0.999
2.5 0.9995
2.6 0.9965
2.7 0.987
2.8 0.9585
2.9 0.928
3 0.8505
3.1 0.7305
3.2 0.5625
3.3 0.398
3.4 0.2412
3.5 0.132125
3.6 0.0595882352941177
3.7 0.0258974358974359
3.8 0.0087719298245614
3.9 0.00245343137254902
4 0.000719942404607631
4.1 0.000157315731573157
4.2 3.05772994129158e-05
4.3 5.42267001426162e-06
4.4 1.11684237028812e-06
4.5 1.65271731673548e-07
4.6 8.18057870019694e-08
};
\addlegendentry{comp-comp}
\addplot [semithick, color0, mark=triangle, mark size=2, mark options={solid}]
table {%
2 1
2.1 1
2.2 1
2.3 1
2.4 0.999
2.5 0.9995
2.6 0.997
2.7 0.989
2.8 0.962
2.9 0.934
3 0.8535
3.1 0.7365
3.2 0.575
3.3 0.41
3.4 0.25
3.5 0.13975
3.6 0.0621764705882353
3.7 0.0275135135135135
3.8 0.00964423076923077
3.9 0.00266489361702128
4 0.000780811232449298
4.1 0.000167000668002672
4.2 3.36938576097577e-05
4.3 6.10832503619183e-06
4.4 1.19522674248123e-06
4.5 1.75722001627741e-07
4.6 8.37819156735019e-08
};
\addlegendentry{comp-comp(uni)}
\addplot [semithick, color1, mark=x, mark size=2, mark options={solid}]
table {%
2 1
2.1 1
2.2 1
2.3 1
2.4 0.999
2.5 0.9995
2.6 0.998
2.7 0.99
2.8 0.963
2.9 0.9375
3 0.857
3.1 0.7485
3.2 0.5875
3.3 0.42
3.4 0.257
3.5 0.145
3.6 0.0667333333333333
3.7 0.0285555555555556
3.8 0.00994059405940594
3.9 0.00294134897360704
4 0.000805152979066023
4.1 0.000178858880343409
4.2 3.48565652340618e-05
4.3 6.38329109722391e-06
4.4 1.21412116306581e-06
4.5 2.19935164318683e-07
4.6 1.08618113391439e-07
};
\addlegendentry{comp(uni)-comp(uni)}
\addplot [semithick, color2, mark=asterisk, mark size=2, mark options={solid}]
table {%
2 1
2.1 1
2.2 1
2.3 1
2.4 0.999
2.5 0.9995
2.6 0.998
2.7 0.993
2.8 0.972
2.9 0.9525
3 0.8825
3.1 0.785
3.2 0.6475
3.3 0.475
3.4 0.3065
3.5 0.18
3.6 0.0840833333333333
3.7 0.0385769230769231
3.8 0.0141408450704225
3.9 0.00403225806451613
4 0.00111968680089485
4.1 0.000236923076923077
4.2 4.5134500812421e-05
4.3 6.93514941779421e-06
4.4 1.01033579718904e-06
4.5 9.338108822103e-08
};
\addlegendentry{min-comp(uni)}
\addplot [semithick, gray, mark=triangle*, mark size=2, mark options={solid}]
table {%
2 1
2.1 1
2.2 1
2.3 1
2.4 1
2.5 0.9995
2.6 1
2.7 0.9965
2.8 0.978
2.9 0.9575
3 0.901
3.1 0.8145
3.2 0.685
3.3 0.5195
3.4 0.343333333333333
3.5 0.2052
3.6 0.1052
3.7 0.0473181818181818
3.8 0.0184727272727273
3.9 0.00564971751412429
4 0.00161550888529887
4.1 0.000362844702467344
4.2 6.79532481652623e-05
4.3 1.09716492583165e-05
4.4 1.71465381676256e-06
4.5 1.50238081367718e-07
4.6 2.39317466585299e-08
};
\addlegendentry{min-lut}
\addplot [semithick, red!75.2941176470588!black, mark=sqaure*, mark size=1.225, mark options={solid}, forget plot]
table {%
2 1
2.1 1
2.2 1
2.3 1
2.4 1
2.5 0.999
2.6 0.9995
2.7 0.9955
2.8 0.9845
2.9 0.9655
3 0.921
3.1 0.8445
3.2 0.7235
3.3 0.5725
3.4 0.382666666666667
3.5 0.2486
3.6 0.1345
3.7 0.0665625
3.8 0.0267631578947368
3.9 0.00896428571428571
4 0.00286571428571429
4.1 0.000704721634954193
4.2 0.000159974404095345
4.3 3.15258511979823e-05
4.4 5.63770951137972e-06
4.5 9.47184250937744e-07
4.6 2.19849342131378e-07
4.7 7.43262327005693e-08
};
\addplot [semithick, color0, mark=triangle, mark size=2, mark options={solid}, forget plot]
table {%
2 1
2.1 1
2.2 1
2.3 1
2.4 1
2.5 0.9995
2.6 0.9995
2.7 0.9945
2.8 0.9845
2.9 0.9715
3 0.9285
3.1 0.859
3.2 0.7355
3.3 0.584
3.4 0.401666666666667
3.5 0.264
3.6 0.143857142857143
3.7 0.0718571428571429
3.8 0.0294117647058824
3.9 0.01009
4 0.00306748466257669
4.1 0.000827129859387924
4.2 0.000197199763360284
4.3 3.69863520360987e-05
4.4 6.5585812477045e-06
4.5 1.11917477411585e-06
4.6 2.14209429499087e-07
4.7 5.3616747488949e-08
};
\addplot [semithick, color1, mark=x, mark size=2, mark options={solid}, forget plot]
table {%
2 1
2.1 1
2.2 1
2.3 1
2.4 1
2.5 0.9995
2.6 0.9995
2.7 0.9965
2.8 0.9845
2.9 0.971
3 0.93
3.1 0.8575
3.2 0.7355
3.3 0.5875
3.4 0.402333333333333
3.5 0.268
3.6 0.145428571428571
3.7 0.0719285714285714
3.8 0.0303939393939394
3.9 0.01008
4 0.00320127795527157
4.1 0.000825082508250825
4.2 0.000187160771102377
4.3 3.58268844941244e-05
4.4 6.31281248421797e-06
4.5 1.01773954690109e-06
4.6 1.98575146945609e-07
};
\addplot [semithick, color2, mark=asterisk, mark size=2, mark options={solid}, forget plot]
table {%
2 1
2.1 1
2.2 1
2.3 1
2.4 1
2.5 0.9995
2.6 0.9995
2.7 0.998
2.8 0.9895
2.9 0.9755
3 0.947
3.1 0.89
3.2 0.789
3.3 0.65
3.4 0.459
3.5 0.3155
3.6 0.187
3.7 0.0911818181818182
3.8 0.042125
3.9 0.0145072463768116
4 0.00455203619909502
4.1 0.00117924528301887
4.2 0.000264620269912675
4.3 4.88495921059059e-05
4.4 7.94382129579613e-06
4.5 1.04977420537062e-06
4.6 1.53805324680025e-07
4.7 2.06288863348363e-08
};
\addplot [semithick, gray, mark=triangle*, mark size=1.75, mark options={solid}, forget plot]
table {%
2 1
2.1 1
2.2 1
2.3 1
2.4 1
2.5 1
2.6 1
2.7 0.9995
2.8 0.9985
2.9 0.9875
3 0.9775
3.1 0.9315
3.2 0.86
3.3 0.7615
3.4 0.591
3.5 0.429
3.6 0.28625
3.7 0.162571428571429
3.8 0.083
3.9 0.0324193548387097
4 0.0124197530864198
4.1 0.00396825396825397
4.2 0.00106716417910448
4.3 0.000241721053903795
4.4 4.70521808685833e-05
4.5 7.54523367588694e-06
4.6 1.204279922573e-06
4.7 1.37685416360699e-07
4.8 3.6144578313253e-08
};
\draw (axis cs:4.05,5e-4) ellipse (0.4cm and 0.2cm);
\node at (axis cs:3.99,3e-4){\scriptsize 4 bits};
\draw (axis cs:4.18,5e-4) ellipse (0.8cm and 0.2cm);
\node at (axis cs:4.29,5e-4){\scriptsize 3 bits};
\end{axis}

\begin{axis}[
axis background/.style={fill=white},
xshift=0.64*\figurewidth,
yshift=0.56*\figureheight,
height=0.4*\figureheight,
width=0.35*\figurewidth,
legend cell align={left},
legend style={
    fill opacity=0.8,
    draw opacity=1,
    text opacity=1,
    at={(0.03,0.03)},
    anchor=south west,
    draw=white!80!black
},
log basis y={10},
tick align=outside,
tick pos=left,
x grid style={white!69.0196078431373!black},
xlabel={},
xmajorgrids,
xmin=3.99, xmax=4.03,
xminorgrids,
minor x tick num=9,
xtick style={color=black},
xtick={3.98,4.0,4.02,4.04},
xticklabels={3.98,4.0,4.02,4.04},
y grid style={white!69.0196078431373!black},
ylabel={},
ymajorgrids,
ymin=5e-04, ymax=0.0009,
yminorgrids,
ymode=log,
ytick style={color=black},
minor y tick num=10,
ytick={0.001,0.0006,0.0001},
yticklabels={0.001,$6\cdot 10^{-4}$,0.0001},
]
\addplot [semithick, red!75.2941176470588!black, mark=square*, mark size=1.225, mark options={solid}]
table {%
    2 1
    2.1 1
    2.2 1
    2.3 1
    2.4 0.999
    2.5 0.9995
    2.6 0.9965
    2.7 0.987
    2.8 0.9585
    2.9 0.928
    3 0.8505
    3.1 0.7305
    3.2 0.5625
    3.3 0.398
    3.4 0.2412
    3.5 0.132125
    3.6 0.0595882352941177
    3.7 0.0258974358974359
    3.8 0.0087719298245614
    3.9 0.00245343137254902
    4 0.000719942404607631
    4.1 0.000157315731573157
    4.2 3.05772994129158e-05
    4.3 5.42267001426162e-06
    4.4 1.11684237028812e-06
    4.5 1.65271731673548e-07
    4.6 8.18057870019694e-08
};
\addplot [semithick, color0, mark=triangle, mark size=2, mark options={solid}]
table {%
    2 1
    2.1 1
    2.2 1
    2.3 1
    2.4 0.999
    2.5 0.9995
    2.6 0.997
    2.7 0.989
    2.8 0.962
    2.9 0.934
    3 0.8535
    3.1 0.7365
    3.2 0.575
    3.3 0.41
    3.4 0.25
    3.5 0.13975
    3.6 0.0621764705882353
    3.7 0.0275135135135135
    3.8 0.00964423076923077
    3.9 0.00266489361702128
    4 0.000780811232449298
    4.1 0.000167000668002672
    4.2 3.36938576097577e-05
    4.3 6.10832503619183e-06
    4.4 1.19522674248123e-06
    4.5 1.75722001627741e-07
    4.6 8.37819156735019e-08
};
\addplot [semithick, color1, mark=x, mark size=2, mark options={solid}]
table {%
    2 1
    2.1 1
    2.2 1
    2.3 1
    2.4 0.999
    2.5 0.9995
    2.6 0.998
    2.7 0.99
    2.8 0.963
    2.9 0.9375
    3 0.857
    3.1 0.7485
    3.2 0.5875
    3.3 0.42
    3.4 0.257
    3.5 0.145
    3.6 0.0667333333333333
    3.7 0.0285555555555556
    3.8 0.00994059405940594
    3.9 0.00294134897360704
    4 0.000805152979066023
    4.1 0.000178858880343409
    4.2 3.48565652340618e-05
    4.3 6.38329109722391e-06
    4.4 1.21412116306581e-06
    4.5 2.19935164318683e-07
    4.6 1.08618113391439e-07
};
\end{axis}
\end{tikzpicture}

    \vspace{-0.6cm}
    \caption{Frame error rates for scenario 1 (3 and 4 bits).} %
    \label{fig:ber_code2_3_4bit}
\end{figure}
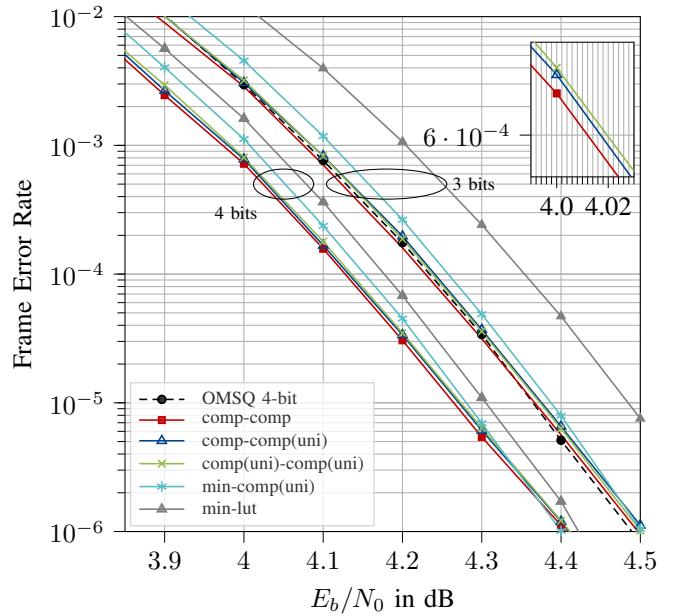

\subsubsection{Scenario 1}\label{sec:scenario1}
We analyze the performance for a rate $R{=}0.841$ code from \cite{noauthor_standard_2006} with $N{=}2048$, $d_c{=}32$ and $d_v{=}6$. First, the evolution of mutual information in Fig.~\ref{fig:evolution_of_mi} is evaluated. 

For the 4-bit decoding in Fig.~\ref{fig:mi_ctv_code2_4bit} the non-uniform quantization configuration in check and variable node converges fastest. Remarkably, only very small degradation occurs, when restricting to uniform quantization in check and/or variable node. Using the minimum approximation in the check node results in more significant mutual information loss.
The lookup table based variable node is significantly outperformed by the computational domain solutions.

Additionally for 3-bit decoding in Fig.~\ref{fig:mi_ctv_code2_3bit}, uniform quantization in check and variable node leads only to minor performance degradation. We note that the loss from using concatenated lookup tables in the variable node is more severe for 3-bit decoding because of the internal quantization effects.

The frame error rate results for a maximum of 10 decoding iterations are depicted in Fig.~\ref{fig:ber_code2_3_4bit}. The mutual information performances differences translate to high frame error rates~${>}10^{-4}$. The performance loss from the uniform compared to the non-uniform configurations is less than~0.01\,dB.
In the region ${<}10^{-5}$ cycle effects start to dominate the performance, which are not taken into account in discrete density evolution. For low error rates the minimum approximation closes the performance gap to the computational domain approach, which indicates more robustness against cycle effects. This phenomenon was also observed in \cite{mohr_coarsely_2021}.
The 3-bit uniform decoders achieve similar performance as the 4-bit conventional offset-min-sum decoder (OMSQ).
\subsubsection{Scenario 2}
A medium-rate $R{=}0.5$ code with $N{=}8000$, $d_c{=}6$ and $d_v{=}3$ is considered, which was also studied in \cite{lewandowsky_information-optimum_2018} with label '8000.4000.465'.
As observed in scenario 1, the bit error rates in Fig.~\ref{fig:ber_code2_3_4bit} confirm that restriction to uniform quantization introduces only barely noticeable performance differences. Especially for 3-bit decoding, the computational domain decoder achieves a gain of 0.08\,dB over the lookup table based configuration. The 3-bit decoder with minimum approximation in the check node looses only 0.04\,dB compared to the 4-bit OMSQ decoder.

\section{Conclusions}
This paper revealed significant complexity reduction potential with minor performance loss for mutual information maximizing decoding by replacing non-uniform with uniform quantization operations. The proposed hardware structure essentially eliminated most of the resources required for non-uniform quantization in the computational domain approach for check and variable nodes.
For the check node update, the minimum approximation still achieved the lowest complexity at the price of performance losses ranging from 0.015-0.025\,dB compared to computational domain with uniform quantization.

The most promising candidate was a 3-bit decoder architecture using the variable node with uniform quantization and check node with minimum approximation. Performance close to 4-bit OMSQ decoding is achieved with the potential to significantly save wiring and register complexity.

\section{Appendix}
\subsection{Computational Domain Check Node Update Derivation}\label{app:derivation_cn_operation}
This section uses results from \cite{he_mutual_2019, stark_machine_2021} to derive the check node update (\ref{equ:cn_update}). For simplicity, consider a check node of degree $d_c{=}3$. Bit $b_3$ can be obtained from the parity check equation with $b_3{=}b_1{\oplus}b_{2}$. Correspondingly, soft information can be computed from the probabilities $p(b_1)$ and $p(b_2)$ through circular convolution $p(b_3){=}\sum_{b_1}p(b_1)p(b_2{=}b_3{\ominus} b_1)$, which is equivalently performed in frequency domain using the DFT as $p(b_3){=}\mathcal{F}^{-1}\left\{\mathcal{F}\{p(b_1)\}\mathcal{F}\{p(b_2)\}\right\}$. Then, $p(b_3{=}0){=}\frac{1}{2}{+}\frac{1}{2}\left(p(b_1{=}0){-}p(b_1{=}1)\right)\left(p(b_2{=}0){-}p(b_2{=}1)\right)$. The result can be generalized for degree $d_c{>}3$, where $p(b_{d_c}{=}0){=}\frac{1}{2}{+}\frac{1}{2}\prod_{i{=}1}^{d_c-1}\left(p(b_i{=}0){-}p(b_i{=}1)\right)$. The corresponding LLR is given by
\begin{align}
\begin{split}
        L(b_{d_c})&=\log\left(\frac{\frac{1}{2}+\frac{1}{2}\prod_{i=1}^{d_c-1}\psi_i}{\frac{1}{2}-\frac{1}{2}\prod_{i=1}^{d_c-1}\psi_i}\right)
        =2\tanh^{-1}\prod_{i=1}^{d_c-1}\psi_i,
\end{split}
\end{align}
where $\psi_i{=} p(b_i=0)-p(b_i=1)=\tanh \frac{L(b_i)}{2}$. Performing the multiplication in log-domain, we get:
\begin{align}
\begin{split}
L(b_{d_c})&=\left(\prod_{i=1}^{d_c-1} \operatorname{sgn}\psi_i\right)2\tanh^{-1}\exp\left(\sum_{i=1}^{d_c-1}\log|\psi_i|\right).
\label{equ:llr_cn_output}
\end{split}
\end{align}
It is shown in \cite{kurkoski_quantization_2014}, that using a binary LLR as the input $y\coloneqq L(b_{d_c})$ of a threshold quantizer $t{=}Q(y)$, mutual information maximizing compression $\max_{Q} I(\myvar{X};\myvar{T})$ is achievable. 
The monotonic increasing $\tanh^{-1}$-function has no effect on the order of input events to the quantizer. Furthermore, note that $\exp(-x)$ is a monotonic decreasing function of $x$ with $x=\sum_{i=1}^{d_c-1}(-\log|\psi_i|)$, which only inverts the order w.r.t. the magnitude of the input events. Thus, threshold quantization is equivalently performed with
\begin{align}
\begin{split}
y\coloneqq\left(\prod_{i=1}^{d_c-1} \operatorname{sgn}(\psi_i)\right)\sum_{i=1}^{d_c-1}(-\log|\psi_i|).
\end{split}
\end{align}

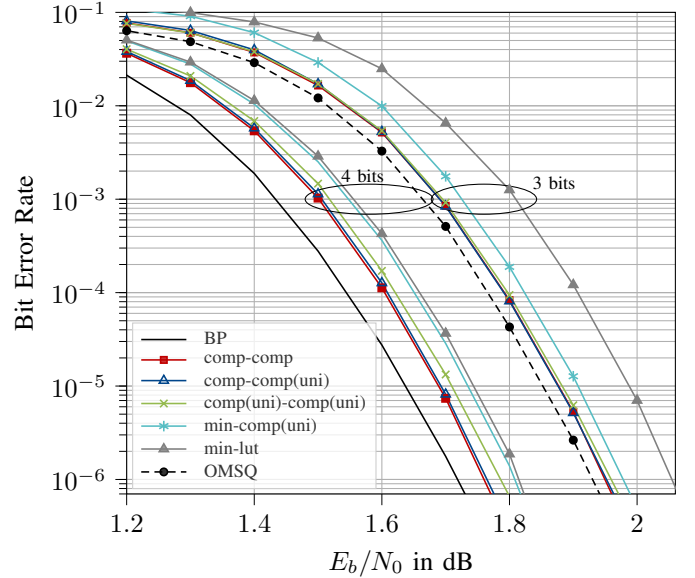
\begin{figure}[t]
    \hspace*{-0.7cm}  
    
    \pgfkeys{/includestandaloneresized, default, width=1.0\linewidth, ratio=0.9}
    \setlength\figurewidth{\itgWidth}
    \setlength\figureheight{\itgRatio\figurewidth}
    \pgfplotsset{legend style={nodes={scale=0.7, transform shape}}}
\pgfplotsset{ every non boxed x axis/.append style={x axis line style=-},
               every non boxed y axis/.append style={y axis line style=-}}
\begin{tikzpicture}

\definecolor{color0}{rgb}{0,0.270588235294118,0.541176470588235}
\definecolor{color1}{rgb}{0.584313725490196,0.729411764705882,0.309803921568627}
\definecolor{color2}{rgb}{0.341176470588235,0.741176470588235,0.764705882352941}

\begin{axis}[
height=\figureheight,
legend cell align={left},
legend style={
  fill opacity=0.8,
  draw opacity=1,
  text opacity=1,
  at={(0.03,0.03)},
  anchor=south west,
  draw=white!80!black,
  at={(0.01,0.01)}
},
log basis y={10},
tick align=outside,
tick pos=left,
width=\figurewidth,
x grid style={white!69.0196078431373!black},
xlabel={\(\displaystyle E_b/N_0\) in dB},
xmajorgrids,
xmin=1.2, xmax=2.06,
xminorgrids,
xtick style={color=black},
y grid style={white!69.0196078431373!black},
ylabel={Bit Error Rate},
ymajorgrids,
ymin=7e-07, ymax=0.1,
yminorgrids,
ymode=log,
ytick style={color=black}
]
\addplot [semithick, black]
table {%
1 0.061351125
1.1 0.042502125
1.2 0.0212653333333333
1.3 0.00798082142857143
1.4 0.0018885625
1.5 0.000278295774647887
1.6 2.74904434250765e-05
1.7 1.76700037778617e-06
1.8 8.31588132635253e-08
};
\addlegendentry{BP}
\addplot [semithick, red!75.2941176470588!black, mark=square*, mark size=1.225, mark options={solid}]
table {%
1 0.09559075
1.1 0.0879795
1.2 0.077043
1.3 0.060577375
1.4 0.037269125
1.5 0.016506625
1.6 0.00518344444444444
1.7 0.000849559782608696
1.8 8.30604938271605e-05
1.9 5.29956122856003e-06
2 1.84016824395373e-07
};
\addlegendentry{comp-comp}
\addplot [semithick, color0, mark=triangle, mark size=2, mark options={solid}]
table {%
1 0.1007045
1.1 0.0932685
1.2 0.081019875
1.3 0.064068625
1.4 0.039667
1.5 0.0171278125
1.6 0.005279625
1.7 0.000840520408163265
1.8 8.12448512585812e-05
1.9 5.17045692726892e-06
2 2.20147763578275e-07
};
\addlegendentry{comp-comp(uni)}
\addplot [semithick, color1, mark=x, mark size=2, mark options={solid}]
table {%
1 0.095224625
1.1 0.0872845
1.2 0.0762115
1.3 0.06030825
1.4 0.037884625
1.5 0.01715325
1.6 0.0053890625
1.7 0.000919552631578947
1.8 9.41610169491525e-05
1.9 6.22306501547988e-06
2 2.86738746690203e-07
};
\addlegendentry{comp(uni)-comp(uni)}
\addplot [semithick, color2, mark=asterisk, mark size=2, mark options={solid}]
table {%
1 0.12514275
1.1 0.119148
1.2 0.108493875
1.3 0.0910945
1.4 0.060709
1.5 0.0290853333333333
1.6 0.00993946875
1.7 0.00175825657894737
1.8 0.000188503174603175
1.9 1.27337607861937e-05
2 4.91319107662464e-07
};
\addlegendentry{min-comp(uni)}
\addplot [semithick, gray, mark=triangle*, mark size=2, mark options={solid}]
table {%
1 0.12422775
1.1 0.118072
1.2 0.11030225
1.3 0.099610375
1.4 0.078949
1.5 0.053189
1.6 0.0249069166666667
1.7 0.006528425
1.8 0.00125910326086957
1.9 0.000121317118226601
2 7.01954277286136e-06
2.1 1.9730127576055e-07
};
\addlegendentry{min-lut}
\addplot [semithick, red!75.2941176470588!black, mark=square*, mark size=1.4, mark options={solid}, forget plot]
table {%
1 0.072753625
1.1 0.05781325
1.2 0.036532875
1.3 0.0176799166666667
1.4 0.00539747222222222
1.5 0.00102787820512821
1.6 0.000112603773584906
1.7 7.39523220665018e-06
1.8 2.64430977814297e-07
};
\addplot [semithick, color0, mark=triangle, mark size=2, mark options={solid}, forget plot]
table {%
1 0.07402025
1.1 0.05928425
1.2 0.038132125
1.3 0.018617
1.4 0.00576490625
1.5 0.00113658783783784
1.6 0.000126987847222222
1.7 8.16306168015629e-06
1.8 3.18431213168055e-07
};
\addplot [semithick, color1, mark=x, mark size=2, mark options={solid}, forget plot]
table {%
1 0.074111375
1.1 0.0604215
1.2 0.040367125
1.3 0.020861
1.4 0.00687221428571429
1.5 0.00146871153846154
1.6 0.000170808333333333
1.7 1.3296221570066e-05
1.8 6.58884131656409e-07
};
\addplot [semithick, color2, mark=asterisk*, mark size=2, mark options={solid}, forget plot]
table {%
1 0.0820325
1.1 0.069194375
1.2 0.0495345
1.3 0.0279738333333333
1.4 0.0104195
1.5 0.00251398611111111
1.6 0.000362799528301887
1.7 2.85849785407725e-05
1.8 1.3952260559191e-06
1.9 2.6141488162345e-08
};
\addplot [semithick, gray, mark=triangle*, mark size=2, mark options={solid}, forget plot]
table {%
1 0.081358125
1.1 0.06923675
1.2 0.050571375
1.3 0.029370375
1.4 0.011383
1.5 0.0028848
1.6 0.000427679411764706
1.7 3.6646714922049e-05
1.8 1.87091194968553e-06
1.9 2.5462962962963e-08
};
\addplot [semithick, black, densely dashed, mark=*, mark size=1.4, mark options={solid}]
table {%
1 0.082112
1.1 0.07486775
1.2 0.06359525
1.3 0.0486055
1.4 0.0289625
1.5 0.0121364375
1.6 0.00327465909090909
1.7 0.000511422131147541
1.8 4.29071936056838e-05
1.9 2.63681411650955e-06
2 1.0619667714414e-07
};
\addlegendentry{OMSQ}
\draw (axis cs:1.58,1e-3) ellipse (0.85cm and 0.2cm);
\node at (axis cs:1.57,1.8e-3){\scriptsize 4 bits};
\draw (axis cs:1.76,1e-3) ellipse (0.7cm and 0.2cm);
\node at (axis cs:1.87,1.5e-3){\scriptsize 3 bits};
\end{axis}

\end{tikzpicture}

    \vspace{-0.6cm}
    \caption{Bit error rates for scenario 2 (3 and 4 bit).}
    \label{fig:ber_code1_3_4bit}
\end{figure}

\bibliographystyle{MyIEEEtran}
\bibliography{literature}
\newpage

\newpage

\end{document}